\documentclass[runningheads]{llncs}

\usepackage{lscape}
\usepackage{rotating}
\usepackage{latexsym}
\usepackage{dcolumn}
\usepackage{booktabs}
\usepackage{amsmath}
\usepackage[algo2e,ruled,linesnumbered,boxed,commentsnumbered]{algorithm2e}
\usepackage{multirow}
\usepackage{graphicx}
\usepackage{amssymb}
\usepackage{pifont}

\usepackage{graphicx}
\begin{document}
\title{Graph-based Ontology Summarization: A Survey}
%
%
\author{Seyedamin Pouriyeh\inst{1} \and
Mehdi Allahyari\inst{2}\and
Qingxia Liu\inst{3}
\and
Gong Cheng\inst{3}\and
\mbox{Hamid Reza Arabnia}\inst{1}\and
Yuzhong Qu\inst{3}\and
Krys Kochut\inst{1}}
\authorrunning{S. Pouriyeh et al.}
%
\institute{Computer Science Department, University of Georgia, Athens, GA, USA.\\
\email{\{pouriyeh, hra, kochut\}@uga.edu}\and
Computer Science Department, Georgia Southern University, Statesboro, GA, USA.\\
\email{mallahyari@georgiasouthern.edu}\\
\and
National Key Laboratory for Novel Software Technology, Nanjing University, Nanjing, China.\\
\email{qxliu.nju@gmail.com, \{gcheng, yzqu\}@nju.edu.cn}}
\maketitle              
\begin{abstract}
Ontologies have been widely used in numerous and varied applications, e.g., to support data modeling, information integration, and knowledge management.
With the increasing size of ontologies, ontology understanding, which is playing an important role in different tasks,
is becoming more difficult. Consequently, ontology summarization, as a way to distill key information from an ontology and generate an abridged version to facilitate a better understanding, is getting growing attention.
In this survey paper, we review existing ontology summarization techniques and focus mainly on graph-based methods, which represent an ontology as a graph and apply centrality-based and other measures to identify the most important elements of an ontology as its summary. After analyzing their strengths and weaknesses, we highlight a few potential directions for future research.

\end{abstract}

\section{Introduction}
An \emph{ontology} provides an explicit specification of a vocabulary for a shared domain~\cite{gruber1993translation}. \emph{Terms} in that vocabulary are mainly classes and properties denoting concepts and their relationships in the domain, respectively, forming a conceptualization of the world that we wish to represent for some purpose. In an ontology, the interpretation and use of terms are constrained by formal \emph{axioms}. As ontologies can help people and organizations reach consensus on conceptualizations, they have found wide application in knowledge management, information integration, data access, etc. In particular, they play an important role in the recent explosive growth of Semantic Web deployment, where an ontology is frequently used as the schema of a knowledge base.

With the dramatic growth in both size and complexity of ontologies, their comprehension, exploration, and exploitation are becoming increasingly difficult. Summarization, in order to generate an overview or a preview of an ontology, is one possible solution that has received increasing research attention, recently. \emph{Ontology summarization} is defined as a technique of distilling key information from an ontology in order to produce an abridged version for different tasks~\cite{zhang2007ontology}. The output is a compact ontology summary, for a better and quicker understanding of an ontology, which can facilitate and reduce the cost of the next tasks in various applications such as ontology evaluation~\cite{brank2005survey}, matching~\cite{om}, and search.

Compared with an early literature review~\cite{li2010ontology}, we have witnessed the emergence of many ontology summarization techniques, in recent years. In this survey paper, rather than providing a comprehensive bibliography, we mainly sort, review, and compare various \emph{graph-based methods} for ontology summarization. An ontology can be transformed into different graph models to represent the relations between terms and/or axioms. A broad range of measures have been presented to assess the importance of each node, which can be a term or an axiom. A subset of top-ranked nodes form an ontology summary, so the output of an ontology summarization approach is usually \emph{a list of ranked terms or axioms}. Some approaches further choose paths to connect selected nodes and return \emph{a subgraph}.

Table~\ref{tb:methods} summarizes the methods that will be reviewed in this paper. We will first compare different graph models, and then discuss measures for assessing node importance including centrality-based, coverage-based, and others. Finally, we conclude the paper with future directions.

Note that our survey focuses on the summarization of terminological definitions in ontologies (i.e., TBox). Methods for summarizing instance data in knowledge bases (i.e., ABox), e.g., \cite{abox}, will not be addressed.

\begin{table*}
\centering
\caption{Ontology Summarization Methods}	
\begin{tabular}{lllllll}
\hline
& \textbf{Output} & \textbf{Graph Model} & \textbf{Centrality} & \textbf{Other Measures} \\ 
\hline 

\cite{zhang2006finding} & ranked terms & vocabulary dependency graph & EC & TC \\ 
 \hline 
\cite{tzitzikas2007ranking} & ranked terms & class graph & DC, BC, EC & - \\ 
 \hline 
\cite{zhang2007ontology} & ranked axioms & RDF sentence graph & DC, BC, EC & Di \\ 
 \hline 
\cite{penin08} & ranked axioms & RDF sentence graph & DC & QR \\ 
 \hline 
\cite{peroni2008identifying} & ranked terms & class graph & DC, PC & Co, NS, Po \\ 
 \hline 
\cite{wu2008identifying} & ranked terms & class graph & EC & - \\ 
 \hline 
\cite{zhang2009summarizing} & ranked axioms & term-sentence graph & EC & Di, Po \\ 
 \hline 
\cite{chen2010semantic} & ranked terms & class graph & EC & - \\ 
 \hline 
\cite{pires2010summarizing} & subgraph & class graph & DC & FC \\ 
 \hline 
\cite{cheng2011biprank} & ranked axioms & term-sentence graph & EC & QR, Ch \\ 
 \hline 
\cite{lera2012ontology} & ranked terms & class graph & DC & - \\ 
 \hline 
\cite{ipm} & subgraph & vocabulary dependency graph & - & QR \\ 
 \hline 
\cite{queiroz2013method} & subgraph & class graph & DC, CC & - \\ 
 \hline 
\cite{troullinou2015rdf} & subgraph & class graph & RC & - \\ 
 \hline 
\cite{butt2016dwrank} & ranked terms & class graph & EC & QR \\ 
 \hline 
\multirow{2}{*}{\cite{pappas2017exploring}} & \multirow{2}{*}{subgraph} & \multirow{2}{*}{class graph} & DC, BC, EgC & \multirow{2}{*}{-} \\
& & & BrC, HC, Ra & \\ 
 \hline 
\cite{troullinou2017ontology} & subgraph & class graph & RC & - \\ 
 \hline 

\end{tabular}
\label{tb:methods}
\end{table*}
\section{Graph Models}\label{sec:models}
An ontology provides definitions (i.e., axioms) for a set of terms. To represent the relations between terms and/or axioms, various graph models have been developed. In this section we review, illustrate, and compare those models.


\subsection{RDF Graph}
\begin{figure}[t]
  \centering
  \includegraphics[scale=.55]{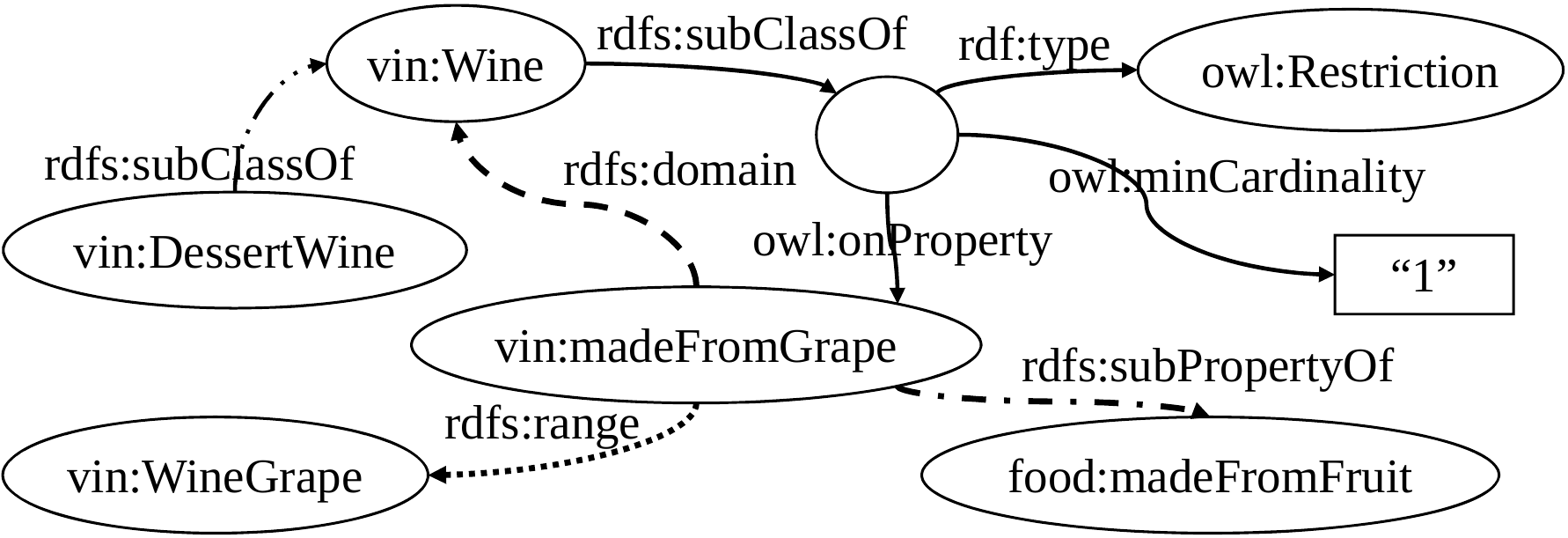}
\caption{An example RDF Graph.}
\label{fig:rdf-g}
\end{figure}

An ontology encoded in RDFS or OWL, which are languages recommended by W3C, can be transformed into an \emph{RDF graph} as illustrated in Fig.~\ref{fig:rdf-g}. Each node-edge-node triple in the graph is called an RDF triple. In this example ontology, three classes and two properties are described by five axioms which are distinguished by different line styles in the figure.

RDFS is an extension of RDF; it is straightforward to represent an RDFS ontology as a graph. In such a graph, all the terms defined in an ontology are represented by nodes. Nodes are connected by directed arcs representing relations between two classes (e.g., \texttt{rdfs:subClassOf}), between two properties (e.g., \texttt{rdfs:subPropertyOf}), or between a property and a class (e.g., \texttt{rdfs:domain}, \texttt{rdfs:range}).

For OWL, W3C provides a document (as part of the OWL language) that defines the mapping of OWL ontologies into RDF graphs. OWL is more expressive than RDFS, and allows complex term definitions. Some axioms, e.g., \texttt{owl:Restriction}, which involves multiple terms, are transformed into multiple RDF triples connected by blank nodes.

\textbf{Comments.} As a ``standard'' graph representation of ontology, RDF graphs have rich tool support. They can be easily processed, stored, queried, and exchanged. However, in many cases an RDF graph representation of an ontology appears unnatural from the semantics point of view.

\subsection{Class Graph}
\begin{figure}[h]
  \centering
  \includegraphics[scale=.55]{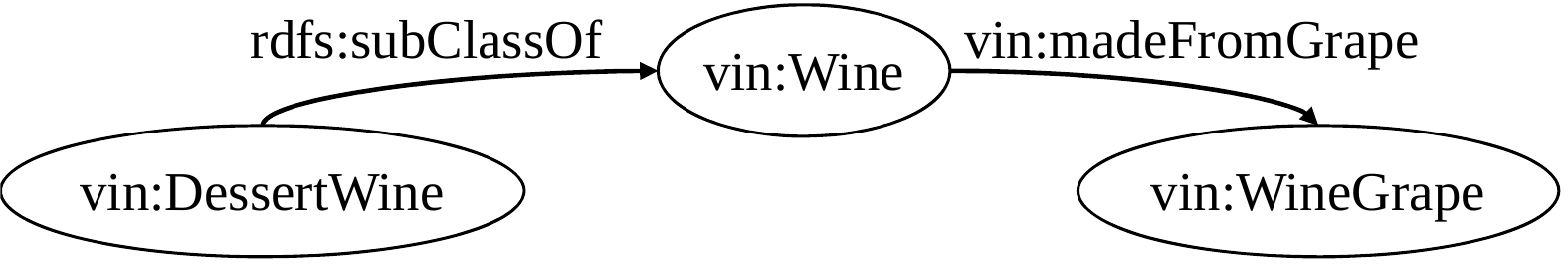}
\caption{An example class graph.}
\label{fig:class-g}
\end{figure}

In order to directly represent semantic relations between classes, Wu \emph{et al.}~\cite{wu2008identifying} presented a graph model where nodes represent classes and directed arcs represent binary relations between classes, which we call a \emph{class graph}. Figure~\ref{fig:class-g} illustrates a class graph for the ontology in Fig.~\ref{fig:rdf-g}. Note that some axioms (e.g., \texttt{owl:Restriction}) are not covered by this graph representation.

As to the relations between classes, if we only allow \texttt{rdfs:subClassOf}, the resulting graph will be a class hierarchy representing subsumption relations, as considered in~\cite{peroni2008identifying}. More generally, a relation can also be a property defined in the ontology, connecting from its domain (which is a class) to its range (also a class).

\textbf{Comments.} Class graphs are close to human cognition. As classes are first-class citizens, class graphs are particularly suitable for approaches to ranking classes. However, the expressivity of class graph is limited. It well supports binary relations between classes but not more complex axioms involving multiple classes, e.g., \texttt{owl:unionOf}.

\subsection{RDF Sentence Graph}
Zhang \emph{et al.}~\cite{zhang2007ontology} proposed an \emph{RDF sentence graph}. An RDF sentence is a subset of RDF triples, and a set of RDF sentences form the finest partition of the triples in an RDF graph such that each blank node only appears in one block. In many cases, an RDF sentence corresponds to an axiom in OWL, since when mapping OWL ontologies into RDF graphs, blank nodes are introduced when an axiom is transformed into multiple RDF triples.

\begin{figure}[h]
  \centering
  \includegraphics[scale=.55]{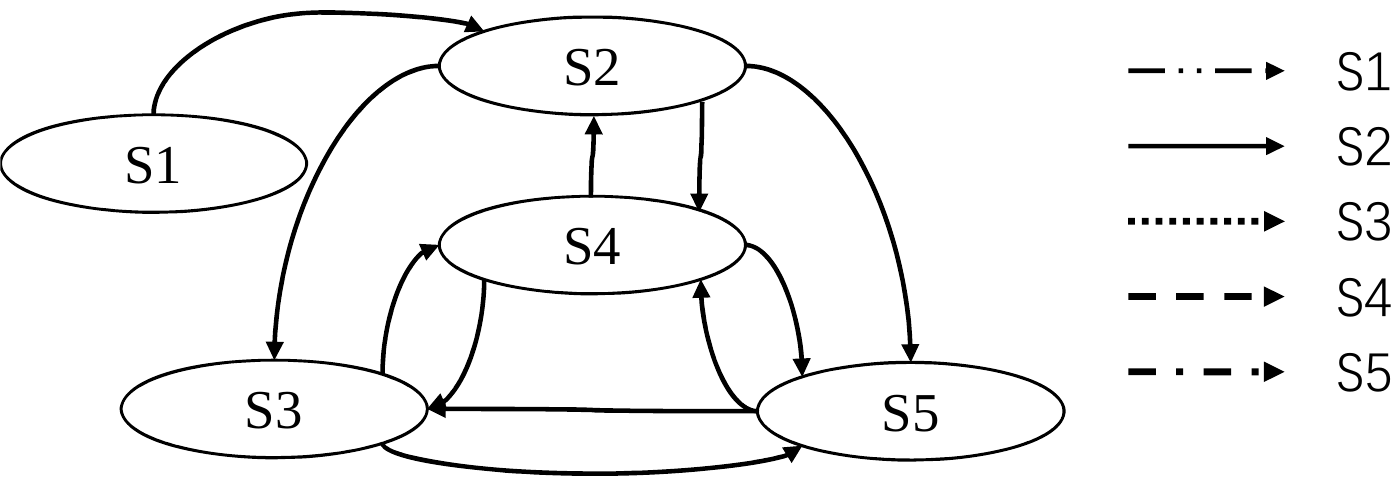}
\caption{An example RDF sentence graph derived from Fig.~\ref{fig:rdf-g}, where each RDF sentence corresponds to a subset of the RDF triples in Fig.~\ref{fig:rdf-g} that have a particular line style.}
\label{fig:rdfsent-g}
\end{figure}

\begin{figure}[h]
  \centering
  \includegraphics[scale=.55]{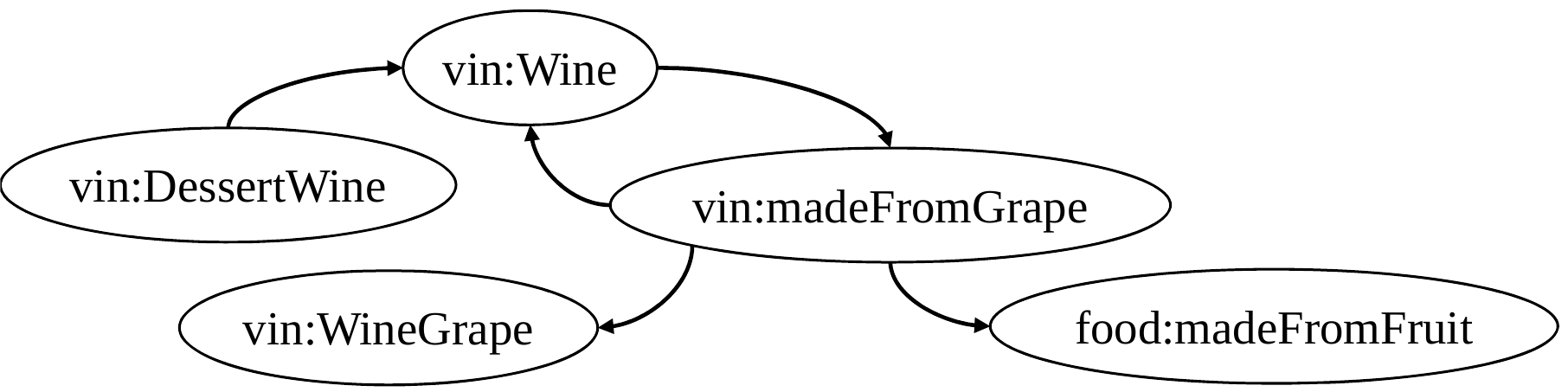}
\caption{An example vocabulary dependency graph.}
\label{fig:vdg}
\end{figure}

In an RDF sentence graph, nodes represent RDF sentences, which are adjacent if the terms they describe overlap. Figure~\ref{fig:rdfsent-g} illustrates an RDF sentence graph for the ontology in Fig.~\ref{fig:rdf-g}; the five RDF sentences exactly correspond to five axioms. Zhang \emph{et al.}~\cite{zhang2007ontology} differentiate between two types of arcs, depending on the structural role of the shared terms, which we will not elaborate. Penin \emph{et al.}~\cite{penin08} further cluster textually similar RDF sentences into topic nodes.

\textbf{Comments.} Compared with RDF triples, there is a better correspondence between RDF sentences and OWL axioms. In an RDF sentence graph, RDF sentences (or roughly speaking, axioms) are first-class citizens, making this model particularly suitable for ranking triples/axioms. However, terms are not explicitly represented in this model, which may limit its application.


\subsection{Vocabulary Dependency Graph}
Based on RDF sentences, Zhang \emph{et al.}~\cite{zhang2006finding} propose \emph{vocabulary dependency graph}, where nodes represent terms, and edges connect terms that co-occur in an RDF sentence. Co-occurrence in an RDF sentence indicates dependency between terms. Figure~\ref{fig:vdg} illustrates a vocabulary dependence graph for the ontology in Fig.~\ref{fig:rdf-g}, derived from Fig.~\ref{fig:rdfsent-g}. Compared with the class graph in Fig.~\ref{fig:class-g}, this new graph covers more terms (e.g., properties), though the edges are unlabeled. Essentially, in a vocabulary dependence graph, each axiom (represented by an RDF sentence) as a complex relation over multiple terms is decomposed into multiple binary relations.

\textbf{Comments.} Compared with the a sentence graph, a vocabulary dependence graph explicitly represents terms in the model, thereby being suitable for ranking terms. Compared with a class graph, a vocabulary dependence graph has both classes and properties as nodes, being suitable for ranking both of them. However, the meaning of an edge in a vocabulary dependence graph is not as explicit as in a class graph.


\subsection{Term-Sentence Graph}
\begin{figure}[h]
  \centering
  \includegraphics[scale=.55]{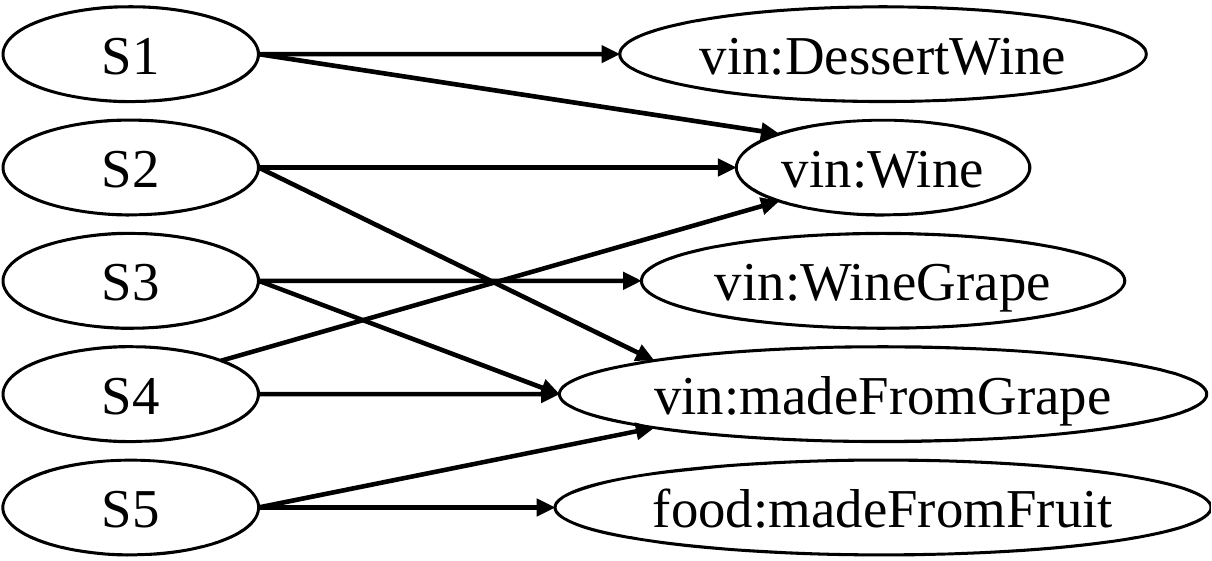}
\caption{An example term-sentence graph.}
\label{fig:term-sent-g}
\end{figure}

Zhang \emph{et al.}~\cite{zhang2009summarizing} present a bipartite graph model, where terms and RDF sentences are both represented by nodes, which we call a \emph{term-sentence graph}. A directed arc connects an RDF sentence to a term if the term is described in that RDF sentence. Figure~\ref{fig:term-sent-g} illustrates a term-sentence graph for the ontology in Fig.~\ref{fig:rdf-g}, derived from Fig.~\ref{fig:rdfsent-g}. Zhang \emph{et al.}~\cite{zhang2009summarizing} differentiate between three types of arcs, depending on the structural role of term in RDF sentence, which we will not elaborate. The model is simplified in~\cite{cheng2011biprank}, where edges are undirected and unlabeled.

\textbf{Comments.} A term-sentence graph is more complex than all the above-mentioned models. One advantage is that, compared with an RDF sentence graph and a vocabulary dependence graph, it explicitly represents both the terms and RDF sentences in the model, thereby expanding its potential application.



\section{Assessment Measures}
In a graph model, a broad range of node importance meanings in the context of ontology summarization has led to many different algorithms. In this section we primarily review popular centrality-based measures. We also discuss coverage-based, application-specific, and other measures.

\subsection{Centrality-based Measures}\label{sect:centrality}
Centrality-based measures are used to find topologically important nodes in a graph representation of an ontology. In general, centrality-based measures are defined via the available structure of the elements of a graph including nodes and edges. These measures primarily focus on the quantitative properties of graph structure such as number of edges and position of nodes, to assess the importance of a node. Some measures take edge types into consideration. As different centrality measures highlight different topological properties of a graph, their outputs are usually not consistent.



\subsubsection{Degree Centrality (DC)}
As one of the simplest centrality measures, \emph{degree centrality} calculates the number of edges incident to a node~$v$:
\begin{equation}\label{eq:dc}
  \text{DC}(v) =  | \text{Number of edges incident to } v | \,.
\end{equation}
\noindent Pappas \emph{et al.}~\cite{pappas2017exploring} use this measure on a class graph to assess the local centrality of each class as its importance. The degree of a class indicates the richness of its description. Nodes with higher degree centrality are more important.

For a directed graph, degree centrality is divided into two categories: \emph{in-degree centrality} and \emph{out-degree centrality}, used in~\cite{zhang2007ontology,queiroz2013method}. The former counts the number of incoming arcs, and the latter counts the number of outgoing arcs.

Instead of considering all the edges incident to~$v$, we may also count only those of specific types. More generally, different types of edges can be assigned different weights, to measure \emph{weighted degree}. For example, Peroni \emph{et al.}~\cite{peroni2008identifying} define the \emph{density} of a class~$v$ as the weighted sum of its number of subclasses, properties, and instances:
\begin{equation}
\begin{split}
  \text{Density}(v) & = w_S * \text{Number of subclasses of } v \\
  & + w_P * \text{Number of properties of } v \\
  & + w_I * \text{Number of instances of } v \,,\\  
\end{split}
\end{equation}
\noindent where $w_S,w_P,w_I$ are weights. Similar methods have been used in~\cite{tzitzikas2007ranking,lera2012ontology}. Pirez \emph{et al.}~\cite{pires2010summarizing} and Queiroz-Sousa \emph{et al.}~\cite{queiroz2013method} divide edges by their types into standard (e.g., is-a, part-of, same-as) and user-defined, which are weighted separately.


\subsubsection*{Relative Cardinality (RC)}
Whereas in the above approaches weights are empirically configured, we highlight \emph{relative cardinality}~\cite{troullinou2015rdf,troullinou2017ontology}, which is a way of automatically weighting edges for calculating weighted degree. In a class graph, the \emph{cardinality} of an edge which represents a property connecting two classes is the number of the corresponding instances of the classes connected with that specific type of property. Therefore, classes and properties having more instances in a knowledge base are considered more important.


\paragraph{Comments on Degree Centrality}
Degree centrality and its variants (e.g., relative cardinality) can be efficiently computed in linear time, which is important when an ontology is very large. However, to assess the importance of a node, these measures mainly use its local information, i.e., the subgraph surrounding that node. Without exploiting the global graph structure, the effectiveness of these measures is limited.


\subsubsection*{Path-based Centrality (PC)}
\emph{Path-based centrality} calculates the number of paths that pass through a particular node. For example, Peroni \emph{et al.}~\cite{peroni2008identifying} count the number of root-leaf paths in a class hierarchy that pass through each class~$v$ as its importance:
\begin{equation}
  \text{PC}(v) = |\text{Number of root-leaf paths passing through } v| \,.
\end{equation}
\noindent A class in the middle of many root-leaf paths is central.


\subsubsection*{Betweenness Centrality (BC)}
As a special case of path-based centrality, it makes sense to only consider \emph{shortest paths}. Specifically, \emph{betweenness centrality}
is defined as the number of shortest paths from all nodes in a graph to all other nodes that pass through that node.
Tzitzikas \emph{et al.}~\cite{tzitzikas2007ranking} use the following implementation of betweenness to assess the importance of each node~$v$ in a class graph:
\begin{equation}\label{eq:bc}
  \text{BC}(v) = \sum\limits_{s\neq v\neq t} \dfrac{ \sigma_{st}(v)  }{ \sigma_{st}  } \,,
\end{equation}
\noindent where $\sigma_{st}$ is the total number of shortest paths from node~$s$ to node~$t$ in the graph, and $\sigma_{st}(v)$ is the total number of those paths passing through node~$v$. The same as degree centrality, a node with a higher betweenness value is considered more important. Betweenness has also been used on RDF sentence graph~\cite{zhang2007ontology}.


\subsubsection*{Ego Centrality (EgC)}
Alternatively, for each node~$v$, let $G_v$ be the subgraph induced by~$v$ and its neighbors, which contains all the edges between them. Pappas \emph{et al.}~\cite{pappas2017exploring} calculate the betweenness centrality of~$v$ within $G_v$, which is called \emph{ego centrality}:
\begin{equation}\label{eq:egc}
  \text{EgC}(v) = \text{BC}(v) \text{ calculated within } G_v \,.
\end{equation}



\subsubsection*{Bridging Centrality (BrC)}
As an improvement to betweenness, Pappas \emph{et al.}~\cite{pappas2017exploring} presented \emph{bridging centrality}. A node with a high bridging centrality is one that connects densely connected components in a graph. To measure that, the bridging centrality of a node~$v$ is defined as the product of $v$'s betweenness centrality ($\text{BC}$) and $v$'s bridging coefficient ($\text{Br}$):
\begin{equation}\label{eq:brc}
\begin{split}
  \text{BrC}(v) & = \text{BC}(v) \cdot \text{Br}(v) \\
  \text{where Br}(v) & =  \dfrac{ \text{DC}(v)^{-1} }{ \sum_{u \in N(v)}{\text{DC}(u)^{-1}} } \,,
\end{split}
\end{equation}
\noindent where $\text{DC}(v)$ is the degree of node~$v$ and $N(v)$ is the set of $v$'s neighbors. Betweenness centrality and bridging coefficient characterize global and local features of a node, respectively.


\paragraph{Comments on Path-based Centrality}
Compared with degree centrality, path-based centrality and its variants (e.g., betweenness centrality, bridging centrality) exploit the global graph structure, going beyond the neighborhood of a node. However, it is computationally expensive to calculate betweenness, which involves calculating the shortest paths between all pairs of nodes in a graph.


\subsubsection*{Closeness Centrality (CC)}
Similar to betweenness, \emph{closeness centrality} is another measure for determining the importance of nodes on a global scale within a graph.
A node is usually considered as a key node if it can quickly interact with all the other nodes in a graph, not only with its immediate neighbors. The closeness of a node~$v$ is originally defined as the average length of the shortest paths between~$v$ and all other nodes in a graph:
\begin{equation}\label{eq:cc}
  \text{CC}(v) =  \dfrac{ n-1 }{ \sum_{u \neq v}{d(v,u)} } \,,
\end{equation}
\noindent where $d(v,u)$ is the distance between~$v$ and~$u$, i.e., the number of edges in the shortest path between them, and $n$~is the number of nodes in the graph.

Closeness centrality is used in~\cite{queiroz2013method},
where an improved implementation for assessing the importance of each class~$v$ in a class graph is proposed:
\begin{equation}
  \text{CC}(v) =  \dfrac{ \sum_{u \neq v}{ \dfrac{score(u)}{d(v,u)} } }{ \sum_{u \neq v}{ \dfrac{1}{d(v,u)} }  } \,,  
\end{equation}
\noindent where $score(u)$ is the importance score of node~$u$ determined by some other measure. This new implementation gives emphasis on the classes that are close to other important classes.


\subsubsection*{Harmonic Centrality (HC)}
We have seen several minor modifications made to the definition of closeness. Pappas \emph{et al.}~\cite{pappas2017exploring} present \emph{harmonic centrality}, in which the average distance is replaced by the harmonic mean of all distances: 
\begin{equation}\label{eq:hc}
  \text{HC}(v) = \dfrac{ 1 }{  \sum_{u \neq v}{d(v,u)} } \,.
\end{equation}


\subsubsection*{Radiality (Ra)}
Pappas \emph{et al.}~\cite{pappas2017exploring} also present \emph{radiality}, which takes the diameter of a graph into account:
\begin{equation}\label{eq:ra}
  \text{Ra}(v) =  \dfrac{ 1 }{ \sum_{u \neq v}{(D - d(v,u)^{-1})} } \,,
\end{equation}
\noindent where $D$~is the diameter of the graph, namely the greatest distance between any pair of nodes in the graph.


\paragraph{Comments on Closeness Centrality}
Closeness centrality and its variants (e.g., harmonic centrality, radiality) are similar to betweenness, also involving calculating the shortest paths between all pairs of nodes in a graph. One difference is that, a node with a high closeness value is usually located at the center of the graph (in terms of distance), but such a node may not have a high betweenness value because it may not be a bridging node that resides in many shortest paths connecting other nodes.




\subsubsection*{Eigenvector Centrality (EC)}
A widely adopted principle is that a node is important if it is connected with important nodes. For example, in a class graph, a class is important if the classes it connects with are important. This gives rise to \emph{eigenvector centrality} which iteratively calculates the importance of each node~$v$ in a graph:
\begin{equation}\label{eq:ec}
  \text{EC}(v) = \dfrac{1}{\lambda} \sum_{u \in N(v)}{\text{EC}(u)} \,,
\end{equation}
\noindent where $N(v)$ is the set of $v$'s neighbors, and $\lambda$~is a constant factor for normalization. The eigenvector centrality of a node is the sum of the eigenvector centrality of its neighbors. The computation iterates over all the nodes in the graph, one round after another until convergence.

Whereas this basic measure has been used in~\cite{chen2010semantic}, its improved variants are more popular in the literature. PageRank, a well-known implementation of eigenvector centrality, is used in~\cite{tzitzikas2007ranking,butt2016dwrank}. Different from the above basic measure, PageRank introduces a damping factor which is added to the centrality. Weighted PageRank, weighted HITS, or their variants are used in~\cite{zhang2006finding,zhang2007ontology,zhang2009summarizing,wu2008identifying,cheng2011biprank}, where centrality is defined as a weighted sum. The weight of an edge between~$v$ and~$u$ indicates the strength of the connection between them; a stronger connection will transport more centrality score from~$u$ to~$v$.

\paragraph{Comments on Eigenvector Centrality}
Eigenvector centrality and its variants (e.g., PageRank, HITS) have shown their effectiveness in many applications. However, they require iterative computation over all the nodes in a graph until convergence, which is time-consuming for large graphs.



\subsubsection*{Empirical Comparison of Centrality-based Measures}
It seems that the effectiveness of a centrality-based measure is related to the graph model, and may also depend on the specific ontology to be summarized as the application and the domain of an ontology provide a guideline in order to select a proper set of measures.

Specifically, according to the experiment results presented in~\cite{tzitzikas2007ranking}, the simple degree centrality (DC) appears more effective than PageRank (i.e., EC) on some class graphs. However, Zhang \emph{et al.}~\cite{zhang2007ontology} report that weighted PageRank (i.e., EC) outperforms degree (i.e., DC) on several RDF sentence graphs; both of them are considerably better than betweenness (i.e., BC). Pappas \emph{et al.}~\cite{pappas2017exploring} find that degree (i.e., DC) and betweenness (i.e., BC, EgC, and BrC) are notably better than closeness (i.e., HC and RA) on a few class graphs.

Unfortunately, we could not draw any reliable conclusions from the current empirical results reported in the literature as they all experiment with a small number of ontologies.


\subsection{Coverage-based Measures}
Top-ranked nodes in a graph representation of an ontology may not form the best ontology summary. For many applications, a good summary is expected to have a good coverage of the contents of an ontology, to form a comprehensive and unbiased overview. Accordingly, the quality of a subset of nodes forming a summary is to be assessed as a whole.

\subsubsection*{Coverage (Co)}
Peroni \emph{et al.}~\cite{peroni2008identifying} propose the \emph{coverage} criterion which aims to show how well the selected set of classes are spread over the whole class hierarchy. For each node~$v$, let $N^+(v)$ be the set of nodes covered by~$v$, including~$v$ and its neighbors, i.e., its subclasses and superclasses in the class hierarchy. The coverage of a set of selected nodes~$V$ is defined as the proportion of nodes in the graph that are covered by~$V$:
\begin{equation}\label{eq:co}
  \text{Co}(V) = \dfrac{|\bigcup_{v \in V}{N^+(v)}|}{n} \,,
\end{equation}
\noindent where $n$~is the number of nodes in the graph.

Further, Peroni \emph{et al.}~\cite{peroni2008identifying} consider an interesting measure called \emph{balance} which is directly related to coverage. It measures how balanced the selected nodes are, i.e., the degree to which each selected node contributes to the overall coverage of the set, which is characterized by standard deviation.


\subsubsection*{Diversity-based Re-ranking (Di)}\label{sect:di}
In~\cite{zhang2007ontology,zhang2009summarizing}, the coverage of a summary is improved by a \emph{re-ranking} step after centrality-based ranking. In these approaches, nodes are iteratively selected to form a summary. In each iteration, the next node to be selected may not be the top-ranked one among the remaining nodes, which will be re-ranked such that a node similar to those selected in previous iterations will be penalized. Specifically, let $score(v)$ be the centrality score of node~$v$, and let $sim(v,u)$ be the similarity between nodes~$v$ and~$u$. Given a set of nodes~$V_s$ which are already selected into the summary and a set of candidate nodes~$V_c$, the next node to be selected from~$V_c$ is
\begin{equation}\label{eq:di}
  \arg\max_{v \in V_c}{(score(v) - \sum_{u \in V_s}{sim(v,u)})} \,.
\end{equation}

Zhang \emph{et al.}~\cite{zhang2007ontology,zhang2009summarizing} use this algorithm to rank RDF sentences, where two RDF sentences are similar if they share terms. The resulting ontology summary is diversified with regard to the terms it contains.

\paragraph{Comments on Coverage-based Measures}
Coverage-based methods complement centrality-based measures, but their current implementations are suboptimal. Coverage in Eq.~(\ref{eq:co}) considers the neighborhood of each node, not taking the global graph structure into account. Diversity-based re-ranking in Eq.~(\ref{eq:di}) has a greedy nature, and may not find the optimum summary in terms of centrality and diversity.

\subsection{Application-specific Measures}
The following two methods are not graph-based but are designed for specific applications.

\subsubsection*{Query Relevance (QR)}
A special kind of ontology summary is a snippet presented in search results pages of an ontology search engine. In this application, terms~\cite{ipm,butt2016dwrank} or RDF sentences~\cite{cheng2011biprank,penin08} that are \emph{relevant to a user query} (e.g., containing query keywords) are prioritized for being presented in a snippet, to show the relevance of an ontology to the user's information needs.

\subsubsection*{Frequency of Correspondences (FC)}
Pires \emph{et al.}~\cite{pires2010summarizing} consider applications where an ontology to be summarized can be an integrated ontology obtained by merging several local ontologies. In that case, an important term in the integrated ontology is one that has a high \emph{frequency of correspondences}, namely it finds correspondences to many classes in local ontologies.

\subsection{Other Measures}
In addition to graph-based and application-specific measures, we briefly review other methods used in the literature.


\subsubsection*{Name Simplicity (NS)}
Peroni \emph{et al.}~\cite{peroni2008identifying} emphasize that \emph{natural categories} or \emph{basic objects} are good representers of an ontology. They propose that a natural category normally has a relatively simple label, and hence they assess the importance of a class by the \emph{simplicity of its name}. A class having compound words in the name will be penalized.

\subsubsection*{Textual Centrality (TC)}
Zhang \emph{et al.}~\cite{zhang2006finding} calculate the \emph{textual centrality} of a term in an ontology. Different from the centrality-based measures discussed in Section~\ref{sect:centrality} which are defined over graph structure, the textual centrality of a term is the similarity between its textual description and the one for the whole ontology.

\subsubsection*{Popularity (Po)}
The wide use of a term on the Web suggests its importance. To measure the \emph{popularity} of a term, Peroni \emph{et al.}~\cite{peroni2008identifying} submit the name of the term as a keyword query to a Web search engine and resort to the number of returned results.
Zhang \emph{et al.}~\cite{zhang2009summarizing} calculate the number of websites hosting RDF documents where the term is instantiated.

\subsubsection*{Cohesion (Ch)}
Cheng \emph{et al.}~\cite{cheng2011biprank} measure the quality of a summary as a whole. Different from diversity-based re-ranking described in Section~\ref{sect:di} which penalizes an ontology summary where RDF sentences share terms, such a summary will be awarded in~\cite{cheng2011biprank} as it exhibits \emph{cohesion}.



\section{Future Directions}
We have investigated different graph models and measures for ontology summarization. We believe that other directions to generate more reliable ontology summaries exist, and we are trying to address some of them to conclude our survey.

Although many algorithms for the ontology summarization problem have been proposed, empirical results reported in the literature suggest that none of them consistently generates the best ontology summary.
In an ideal case, the ontology summarization technique needs to be more flexible in the way that users or applications are able to \emph{tune} the model in order to generate different summaries based on different requirements or inputs. In other words, \emph{dynamic or adaptive ontology summarization} can be viewed as an interesting topic to explore.

Defining \emph{new measures}, either graph-based or not,
is another research activity in the context of ontology summarization. 
Ideas may come from thorough investigations into human-made ``gold-standard'' summaries.
Research advances in the field of information retrieval and text summarization, as well as recent research on entity summarization (e.g., \cite{amin} \cite{pouriyeh2018combining}) which is closely related to ontology summarization, can also provide inspiration. In particular, machine learning techniques have not been extensively used for ontology summarization.

The available approaches apply \emph{extractive techniques} to generate the final summary. In the extractive scenario, a subset of the terms and/or axioms from the original input ontology are selected as a summary. \emph{Non-extractive or abstractive} ontology summarization will be a new direction in this area. In that scenario, the key research question is how to define the output of ontology summarization, e.g., as some kind of high-level aggregate representation of terms and axioms.

There is a lack of \emph{evaluation efforts}. To the best of our knowledge, experiments presented in the literature are all based on a small number of ontologies. No benchmark for ontology summarization is available so far.

Dozens of software systems, libraries, or APIs for text summarization are available, many of which are open-source. By comparison, it is rare to see any \emph{software tool support} for summarizing ontologies. In fact, if such a tool or an application aims to directly serve ordinary users, it needs to also address the presentation (e.g., verbalization, visualization) of and the interaction with ontologies, in which some other challenges would emerge. 

Last but not least, almost all of the methods we have discussed generate ontology summaries to be presented to human users. Summaries may also facilitate computer processing in certain tasks.
It would be interesting to explore applications of this kind.

%
%
%
%
\bibliographystyle{splncs04}
 \bibliography{mybib}
\end{document}